\documentstyle[12pt]{article}

\def\hybrid{\topmargin -20pt    \oddsidemargin 0pt
        \headheight 0pt \headsep 0pt
        \textwidth 6.25in       
        \textheight 9.5in       
        \marginparwidth .875in
        \parskip 5pt plus 1pt   \jot = 1.5ex}

\hybrid

\def\baselinestretch{1.2}

\catcode`\@=11

\def\marginnote#1{}
\newcount\hour
\newcount\minute
\newtoks\amorpm
\hour=\time\divide\hour by60
\minute=\time{\multiply\hour by60 \global\advance\minute by-\hour}
\edef\standardtime{{\ifnum\hour<12 \global\amorpm={am}%
        \else\global\amorpm={pm}\advance\hour by-12 \fi
        \ifnum\hour=0 \hour=12 \fi
        \number\hour:\ifnum\minute<10 0\fi\number\minute\the\amorpm}}
\edef\militarytime{\number\hour:\ifnum\minute<10 0\fi\number\minute}

\def\draftlabel#1{{\@bsphack\if@filesw {\let\thepage\relax
   \xdef\@gtempa{\write\@auxout{\string
      \newlabel{#1}{{\@currentlabel}{\thepage}}}}}\@gtempa
   \if@nobreak \ifvmode\nobreak\fi\fi\fi\@esphack}
        \gdef\@eqnlabel{#1}}
\def\@eqnlabel{}
\def\@vacuum{}
\def\draftmarginnote#1{\marginpar{\raggedright\scriptsize\tt#1}}

\def\draft{\oddsidemargin -.5truein
        \def\@oddfoot{\sl preliminary draft \hfil
        \rm\thepage\hfil\sl\today\quad\militarytime}
        \let\@evenfoot\@oddfoot \overfullrule 3pt
        \let\label=\draftlabel
        \let\marginnote=\draftmarginnote
   \def\@eqnnum{(\theequation)\rlap{\kern\marginparsep\tt\@eqnlabel}%
\global\let\@eqnlabel\@vacuum}  }

\def\preprint{\twocolumn\sloppy\flushbottom\parindent 2em
        \leftmargini 2em\leftmarginv .5em\leftmarginvi .5em
        \oddsidemargin -.5in    \evensidemargin -.5in
        \columnsep .4in \footheight 0pt
        \textwidth 10.in        \topmargin  -.4in
        \headheight 12pt \topskip .4in
        \textheight 6.9in \footskip 0pt
        \def\@oddhead{\thepage\hfil\addtocounter{page}{1}\thepage}
        \let\@evenhead\@oddhead \def\@oddfoot{} \def\@evenfoot{} }

\def\numberbysection{\@addtoreset{equation}{section}
        \def\theequation{\thesection.\arabic{equation}}}

\def\underline#1{\relax\ifmmode\@@underline#1\else
        $\@@underline{\hbox{#1}}$\relax\fi}

\def\titlepage{\@restonecolfalse\if@twocolumn\@restonecoltrue\onecolumn
     \else \newpage \fi \thispagestyle{empty}\c@page\z@
        \def\thefootnote{\fnsymbol{footnote}} }

\def\endtitlepage{\if@restonecol\twocolumn \else \newpage \fi
        \def\thefootnote{\arabic{footnote}}
        \setcounter{footnote}{0}}  

\catcode`@=12
\relax

\def\figcap{\section*{Figure Captions\markboth
        {FIGURECAPTIONS}{FIGURECAPTIONS}}\list
        {Figure \arabic{enumi}:\hfill}{\settowidth\labelwidth{Figure
999:}
        \leftmargin\labelwidth
        \advance\leftmargin\labelsep\usecounter{enumi}}}
 \relax
\def\tablecap{\section*{Table Captions\markboth
        {TABLECAPTIONS}{TABLECAPTIONS}}\list
        {Table \arabic{enumi}:\hfill}{\settowidth\labelwidth{Table
999:}
        \leftmargin\labelwidth
        \advance\leftmargin\labelsep\usecounter{enumi}}}
 \relax
\def\reflist{\section*{References\markboth
        {REFLIST}{REFLIST}}\list
        {[\arabic{enumi}]\hfill}{\settowidth\labelwidth{[999]}
        \leftmargin\labelwidth
        \advance\leftmargin\labelsep\usecounter{enumi}}}
 \relax
\makeatletter
\newcounter{pubctr}
\def\publist{\@ifnextchar[{\@publist}{\@@publist}}
\def\@publist[#1]{\list
        {[\arabic{pubctr}]\hfill}{\settowidth\labelwidth{[999]}
        \leftmargin\labelwidth
        \advance\leftmargin\labelsep
        \@nmbrlisttrue\def\@listctr{pubctr}
        \setcounter{pubctr}{#1}\addtocounter{pubctr}{-1}}}
\def\@@publist{\list
        {[\arabic{pubctr}]\hfill}{\settowidth\labelwidth{[999]}
        \leftmargin\labelwidth
        \advance\leftmargin\labelsep
        \@nmbrlisttrue\def\@listctr{pubctr}}}
 \relax
\makeatother
\newskip\humongous \humongous=0pt plus 1000pt minus 1000pt

\newif\ifdtup

\relax

\def\be{\begin{equation}}
\def\ee{\end{equation}}
\def\ba{\begin{eqnarray}}
\def\ea{\end{eqnarray}}

\begin{document}

\renewcommand{\theequation}{\arabic{equation}}

\newcommand{\beq}{\begin{equation}}
\newcommand{\eeq}[1]{\label{#1}\end{equation}}
\newcommand{\ber}{\begin{eqnarray}}
\newcommand{\eer}[1]{\label{#1}\end{eqnarray}}
\newcommand{\eqn}[1]{(\ref{#1})}
\begin{titlepage}
\begin{center}

\hfill hep--th/0702034\\
\vskip -.1 cm
\hfill February 2007\\

\vskip .8in

{\large \bf Renormalization group equations and geometric flows}\footnote{Based
on lectures given at the {\em 5th International School and Workshop on Quantum
Field Theory and Hamiltonian Systems} held in Calimanesti-Caciulata, Romania,
20-26 May 2006; published in the
proceedings Annals of the University of Craiova Physics AUC,
vol. 16, part I, 2006 (eds. R. Constantinescu and S.-O. Saliu). Also based on talks
delivered at 36-\`eme Institut d'\'et\'e {\em Interactions Fondamentales et la
Structure de l'espace-temps}, 14-25 August 2006, Paris, France; {\em International
School on Modern Trends in Mathematical Physics}, 24-30 September 2006, Sofia,
Bulgaria; {\em Workshop on Geometric and Renormalization Group Flows}, 22-24
November 2006, Golm, Germany.}

\vskip 0.8in

{\bf Ioannis Bakas}
\vskip 0.2in
{\em Department of Physics, University of Patras\\
GR-26500 Patras, Greece\\
\footnotesize{\tt bakas@ajax.physics.upatras.gr}}\\

\end{center}

\vskip .8in

\centerline{\bf Abstract}

\noindent
The quantum field theory of two-dimensional sigma models with bulk and
boundary couplings provides a natural framework to realize and unite different
species of geometric flows that are of current interest in mathematics. In
particular, the bulk renormalization group equation gives rise to the Ricci
flow of target space metrics, to lowest order in perturbation theory, whereas
the boundary renormalization group equation gives rise to the mean curvature
flow for embedded branes. Together they form a coupled system of parabolic
non-linear second order differential equations that can be further generalized
to include non-trivial fluxes. Some closely related higher order curvature
flows, such as the Calabi and Willmore flows associated to quadratic
curvature functionals, are also briefly discussed as
they arise in physics and mathematics. However, there is no known
interpretation for them, as yet, in the context of quantum field theory.
\vfill
\end{titlepage}
\eject

\def\baselinestretch{1.2}
\baselineskip 16 pt
\noindent

Quantum field theory provides the main framework to study a large variety of
problems in modern theoretical physics ranging from high energy physics to
condensed matter physics.
At the same time it serves as laboratory for the fruitful exchange of ideas
and techniques between physics and mathematics. Geometric evolution
equations are no exception to this interplay, since, as it turns out,
certain curvature flows are naturally realized
by the renormalization group equations of two-dimensional sigma models.
The main purpose of these notes is to review the emergence of two
distinct geometric flows, namely the Ricci and the mean curvature flows,
as equations for the bulk and boundary couplings of two-dimensional
Dirichlet sigma models that depend on the energy (or length) scale of the
quantum theory. Fixed points of these flows can be reached by imposing
conformal invariance of the bulk and boundary interactions, respectively,
but, in general, there can be trajectories that extend from the
ultra-violet to the infra-red regime of the underlying quantum field
theory. A few other examples of geometric flows will also be discussed
at the end.

Sigma models have all the necessary ingredients to allow for field
theoretic explorations of problems in differential geometry. In their
simplest form they are defined by the classical action
\be
S = {1 \over 4\pi \alpha^{\prime}}
\int_{\Sigma} d^2 z ~ \sqrt{{\det} \gamma} ~ \gamma^{ab}
G_{\mu \nu} (X) \partial_a X^{\mu} \partial_b X^{\nu} ~,
\ee
where $X^{\mu}$ are local coordinates on a Riemannian manifold ${\cal M}$
with metric $G_{\mu \nu}(X)$, called target, and $z^a$ are local coordinates
on the base space manifold $\Sigma$, also called world-sheet, with
corresponding metric $\gamma_{ab}$. Thus, the fields $X^{\mu}$ are maps
from $\Sigma$ to ${\cal M}$ and they are harmonic when the classical
equations of motion are satisfied. The target space metric can be
severely constrained by imposing additional restrictions on the
two-dimensional field theory defined by $S$, which effectively turn
non-linear sigma models into a valuable tool for studying problems
of differential geometry. For example, supersymmetry
of the classical action can be consistently implemented provided
that ${\cal M}$ is K\"ahler manifold and
generalizations thereof, \cite{susy}.
Another example, which is closer related to the type of problems
that will be considered next comes from the requirement of
conformal invariance at
the quantum level. Although the classical sigma model is always
conformal, as can be readily seen by the (in)dependence of its action
upon $\gamma$, conformal invariance at 1-loop is only achieved
when the target space metric is Ricci flat, i.e., $R_{\mu \nu} = 0$.
This simple fact provides the basis for the sigma model description of
string theory in terms of two-dimensional conformal field theories.
It also serves as starting point for the sigma model description
of certain geometric flows by relaxing conformal invariance of the
quantum theory.

Two-dimensional sigma models were used before, in several occasions, as
toy models for non-linear interactions classically as well as quantum
mechanically. It is quite remarkable that they have many properties
in common with four-dimensional Yang-Mills theories including asymptotic
freedom, \cite{polya}, and non-trivial vacuum structure
due to instantons, \cite{askole}. In this
context, the physically interesting cases are described by
non-linear sigma
models with compact target spaces of positive curvature so that
their beta function is negative and the theory becomes free at
very high energies. Thus, in the ultra-violet regime, perturbation
theory can be used reliably to extract information about physical
processes as in non-abelian gauge theories, \cite{gross}. This
particular property of strong interactions had far reaching
consequences for our current understanding of the physical world
({\em Nobel Prize} 2004) and
at the same time motivated the systematic study of renormalization
group flows in simpler non-linear theories.
Two-dimensional sigma models are perturbatively
renormalizable quantum field theories and the scale dependence of their
couplings can be computed order by order in perturbation theory.
This, in effect, induces deformations of their target space metric
with respect to the renormalization group time $t$, given by the
logarithm of world-sheet length scale, which can be formulated
and studied systematically in all generality. The renormalization
of the metric, $G_{\mu \nu} (X; t)$, viewed as generalized coupling,
takes the following form to 1-loop, \cite{fried},
\be
{\partial \over \partial t} G_{\mu \nu} = - \beta (G_{\mu \nu}) =
- R_{\mu \nu}
\ee
and yields the Ricci-flatness condition at the conformal fixed points,
as noted before. This result is only valid at regions of weak
curvature where all higher loop corrections can be dropped
consistently.

The renormalization group equation of the target space metric is no
other but the {\em Ricci flow} which arose independently in
mathematics, \cite{hami1}; see also the recent textbook \cite{chow}
for detailed account of the technical aspects and an extensive
list of references to the original mathematical papers. The Ricci
flow was introduced as tool to address a variety of non-linear problems
in differential geometry and, in particular, the uniformization of
compact Riemannian manifolds. Under some general conditions, appropriate
rescaling, and surgery, when it is necessary, solutions exist and
tend to a constant curvature metric on ${\cal M}$. Fixed points with
constant curvature metrics occur naturally in a variant of the Ricci
flow that follows by suitable rescaling and time reparametrization so
that the overall volume of space remains fixed through out the
evolution. The resulting normalized flow is equivalent to the Ricci
flow but expressed in different variables. This approach has
been successfully implemented in three dimensions leading to a complete
proof of Poincar\'e's conjecture, \cite{perel} ({\em Fields Medal}
2006). Entropy functionals, some of which
have their origin in known monotonicity formulae of two-dimensional
quantum field theory, such as decreasing $c$-function,
\cite{zamo1}, play
important role in the whole subject. The Ricci flow, which is the
simplest example of {\em intrinsic} curvature flows, in more general
mathematical context, provides a non-linear generalization of the
heat equation. It is driven by the intrinsic curvature tensor and
exhibits dissipative properties that wash
away any deviations of the metric from its canonical
(constant curvature) form in a continuous fashion.

The Ricci flow defines a dynamical system in
superspace, which is an infinite dimensional space consisting of
all possible metrics on ${\cal M}$. It can be described as
gradient flow of the Einstein-Hilbert action with respect to the
DeWitt metric in superspace. Since metrics on ${\cal M}$
serve as generalized couplings of the non-linear sigma model,
the identification of the Ricci flow
with the renormalization group equation
on the space of couplings seems natural. It also provides
physical interpretation to the deformation variable $t$ that
otherwise appears adhoc in mathematics. Note, however, a difference
of philosophy between physics and mathematics regarding the
applicability of the Ricci flow. In high energy physics one is
typically interested in
local quantum field theories that exist at very short distances
and so it is more important to obtain valid solutions by integrating
the renormalization group flow backward in $t$.
They correspond to ancient solutions of the Ricci flow, in the
mathematical terminology, that exist at $t = -\infty$ and evolve
forward in time until the formation of singularities. Said
differently, ancient solutions yield unlimited trajectories
of the backward time evolution without running to
singularities, and, hence, two-dimensional
quantum field theories that are asymptotically free in the
ultra-violet regime. In mathematics, on the other hand, one is
only interested in the forward evolution, starting from a given
initial configuration $G_{\mu \nu} (X; t=0)$, regardless
its past existence, and study its
behavior after sufficiently long time. Although the short time existence
of forward solutions is always guaranteed by the parabolic nature
of the flow, there are several intricate things that may happen
further ahead, such as the formation of singularities.
It is precisely here that all mathematical
studies concentrated in recent years and made the Ricci flow into an
effective tool for studying the geometrization problem of manifolds
in low dimensions.

Close to the singularities higher order curvature terms become
important and introduce modifications of the Ricci flow as
predicted by the higher loop perturbative corrections to the
metric beta function. We will not discuss them at all nor
try to incorporate the
effect of non-perturbative corrections to the beta
function due to instanton. Both can have dramatic effect on the
resolution of singularities that may otherwise arise
and they certainly
deserve proper mathematical attention. Here, we only refer to
another way to (partially) avoid the formation of singularities via
the well known mechanism of flux stabilization of
collapsing cycles. This possibility arises
perturbatively
by introducing an anti-symmetric tensor field $B_{\mu \nu} (X)$,
together with the metric in target space, so that the sigma model
action generalizes to
\be
S = {1 \over 4\pi \alpha^{\prime}}
\int_{\Sigma} d^2 z ~ \sqrt{{\det} \gamma} \left( \gamma^{ab}
G_{\mu \nu} (X) - i \epsilon^{ab} B_{\mu \nu} (X) \right)
\partial_a X^{\mu} \partial_b X^{\nu} ~.
\ee

The 1-loop beta functions of this theory form the following coupled
system of
evolution equations, \cite{callan},
\ba
& & {\partial \over \partial t} G_{\mu \nu} = - \beta (G_{\mu \nu}) =
- R_{\mu \nu} + {1 \over 4} H_{\mu \kappa \lambda} {H_{\nu}}^{\kappa
\lambda} ~, \\
& & {\partial \over \partial t} B_{\mu \nu} = - \beta (B_{\mu \nu}) =
{1 \over 2} \nabla_{\lambda} {H^{\lambda}}_{\mu \nu} ~,
\ea
generalizing the Ricci flow. The field strength of the anti-symmetric
tensor field, $H = dB$, enters quadratically in the beta function of the
metric and can balance the shrinking effect of positively curved spaces.
Then, fixed points arise by imposing the conditions
\be
R_{\mu \nu} = {1 \over 4} H_{\mu \kappa \lambda} {H_{\nu}}^{\kappa
\lambda} ~, ~~~~~
\nabla_{\lambda} {H^{\lambda}}_{\mu \nu} = 0
\ee
and correspond to non-trivial conformal field theories (see, also,
\cite{zachos}). A prime example is provided by the $SU(2)$
Wess-Zumino-Witten model, \cite{eddy},
that represents a round 3-sphere stabilized
by fluxes. Thus, singularities that otherwise seem inevitable
on pure metric backgrounds can be
avoided this way. Such generalizations have not been
investigated at all in mathematics and they certainly deserve further
study. The dilaton field $\Phi(X)$
can also be included in the above equations
by adding reparametrizations generated by a gradient vector field
$\xi_{\mu} = - \partial_{\mu} \Phi$.
For review of the subject see, for instance,
\cite{tsey}.

Two-dimensional sigma models can also be used to explore problems
associated with the geometry of submanifolds ${\cal N}$ embedded in
their target space ${\cal M}$. The key point here is to consider
world-sheets with boundary, which for all practical purposes are
taken to have the topology of a disc so that $\partial \Sigma = S^1$.
Then, by imposing Dirichlet boundary conditions on (some of) the sigma
models fields,
\be
X^{\mu} \mid_{\partial \Sigma} = f^{\mu} (y^A) ~,
\ee
amounts to introducing branes in ${\cal M}$ as embedded submanifolds
with local coordinates $y^A$. The embedding functions can be arbitrary,
but they are fixed once and for all in the classical theory.
Equivalently, one
can think of Dirichlet sigma models as two-dimensional field theories
with generalized bulk and boundary couplings,
\be
S = {1 \over 4\pi \alpha^{\prime}}
\int_{\Sigma} d^2 z ~ \sqrt{{\det} \gamma} \gamma^{ab}
G_{\mu \nu}(X) \partial_a X^{\mu} \partial_b X^{\nu} + {1 \over 2\pi
\alpha^{\prime}} \oint_{\partial \Sigma} d \tau G_{\mu \nu}(X)
V^{\mu} \partial_n X^{\nu} ~,
\ee
where $\tau$ is the parameter along the world-sheet boundary and
$\partial_n$ is the derivative operator normal to it. Variation of
this classical action leads to the following set of compatible
boundary conditions,
\be
f_{,A}^{\mu} G_{\mu \nu} \partial_n X^{\nu} = 0 ~, ~~~~~
V^{\mu} = 0 ~.
\ee
Usual $D$-branes are {\em minimal submanifolds} whose extrinsic curvature
vanishes and correspond to conformal invariant boundary conditions
in the quantum theory.

Note, however, that the embedding functions may depend on the energy
scale of the corresponding quantum field theory, as $f^{\mu} (y^A; t)$.
This, in turn, induces a non-vanishing beta function for the boundary
coupling $V^{\mu}$, which, in principle,
can be calculated to all generality order by
order in perturbation theory. The boundary renormalization group
equation is driven by the extrinsic curvature of the brane embedded
in ${\cal M}$ and the 1-loop result yields the deformation, \cite{leigh},
\be
{\partial \over \partial t} f^{\mu} = \sum_{\sigma = 1}^{\rm codim}
H^{\sigma}
{\hat{n}}_{\sigma}^{\mu}
\ee
perpendicular to the brane. The right-hand side is the mean curvature
vector inward to the brane, which is defined in terms of the extrinsic
curvature tensor
\be
K_{AB}^{\sigma} = G_{\mu \nu} {\hat{n}}_{\sigma}^{\mu}
\left(D_A D_B f^{\nu} + \Gamma_{\rho \lambda}^{\nu} f_{,A}^{\rho}
f_{,B}^{\lambda} \right) ,
\ee
using the trace $H^{\sigma} = g^{AB} K_{AB}^{\sigma}$.
The induced metric on the
brane is defined, as usual, by $g_{AB}(y) = G_{\mu \nu} f_{,A}^{\mu}
f_{,B}^{\nu}$ and ${\hat{n}}_{\sigma}^{\mu}$ form a system
of mutually orthogonal
unit normal vectors labeled by the codimension of the brane.
The fixed points correspond to $D$-branes enjoying conformal
boundary conditions irrespective of the bulk conformal invariance.

As discussed in the paper \cite{sourd},
the boundary renormalization group equation for $f^{\mu} (y; t)$
coincides with the so called
{\em mean curvature flow},
which was introduced independently
in mathematics; for extensive reviews see, for instance,
the textbooks \cite{zhu}, and references
therein. It is the prime example of another class of geometric
evolution equations, the {\em extrinsic} curvature flows,
which attracted a lot of attention in recent years. The arena for
dynamics is now provided by
the infinite dimensional space of all possible
embedding functions of a submanifold in Riemannian spaces.
The simplest example is described by the linear theory of two
free bosons, represented by the plane, in which curves can
deform by their extrinsic curvature. Then, the mean curvature
flow is also known as curve shortening flow. Closed
curves have the tendency to shrink, whereas open ones tend to
stretch until they become geodesic lines, \cite{grayson}.
Similar considerations apply to higher dimensions
and to hypersurfaces of arbitrary codimension satisfying
appropriate technical conditions. Branes can also be immersed allowing
for self-intersections. In all cases, the mean curvature flow favors
the dissipation of extrinsic curvature until an equilibrium
configuration is reach or a singularity. There are also
entropy functionals that help us understand the structure
of singularities, \cite{huisk}, but their physical interpretation
is investigated less than the Ricci flow.

When the ambient space is flat or a more general fixed point of
the Ricci flow, the mean curvature flow is defined with respect to
a fixed background metric. However, Dirichlet sigma models
realize the possibility to have the combined system of Ricci and
mean curvature flows and examine competing effects of shrinking
curves on deforming backgrounds. Of course, the deformations
of the target space metric are insensitive to the presence of
embedded branes, while the deformations of branes depend on
the metric through their extrinsic curvature. Such equations have
not been really studied in
mathematics and they are left open to future work.

Further generalizations include the effect of fluxes. An
anti-symmetric tensor field coupling in target space can affect
the boundary renormalization group equation when it has
non-trivial components
induced on the brane. There can also be additional $U(1)$
gauge fields coupled to the world-sheet boundary that are
quite familiar from the theory of open strings and have their
own beta function. Thus, taking into consideration all possible
bulk and boundary couplings, one arrives at a generalized system of
evolution equations. The bulk flows will be the
standard renormalization
group equations for the metric $G_{\mu \nu}$, the anti-symmetric
tensor field $B_{\mu \nu}$ and the dilaton $\Phi$,
if all are present, and they do not depend on boundary
conditions. The boundary flows follow by variation of the so called
Dirac-Born-Infeld action on the submanifold ${\cal N}$, as gradient
flows,
\be
S_{\rm DBI} = \int_{\cal N} d^n y ~ e^{-\Phi}
\sqrt{{\rm det}(g + b + F)} ~,
\ee
where  $g_{AB}$ and $b_{AB}$ are the induced components of the
metric and anti-symmetric tensor field on the brane and $F_{AB}$
is the field strength of the $U(1)$ gauge field living on it.
Specializing to pure metric backgrounds, one recovers the well
known description of the mean curvature flow as gradient flow of
the area functional of the hypersurface ${\cal N}$ so that
minimal submanifolds sit at its extremal (critical) points.
More generally, the equations that result from the action
$S_{\rm DBI}$ lead
to interesting extensions of the basic system of
flows in sigma models, but they will not be discussed any further
in these notes.
We refer the interested reader to the original work \cite{leigh}
(but see also \cite{dorn} for some technical problems) and the
publication \cite{sourd}.

There are several simple solutions of the Ricci and mean curvature
flows that help us understand how geometric objects can deform.
They are designed
to depend on a few moduli that satisfy a truncated system of
simpler evolution equations, and,
as such, they reduce the general problem to
mini-superspace models. Here, we will only present solutions that
depend on one moduli, namely the radius of uniformly contracting
spheres, and refer the interested reader to the literature for
more examples and further details
(see, for instance, \cite{sourd}, and references therein). Note that
spherically symmetric solutions exist for both Ricci and mean curvature
flows because the time evolution will respect all isometries,
if they are initially present. We also
introduce the important concept of solitonic solutions for geometric
flows and mention a few examples in two dimensions.

First, let us consider a round $n$-sphere of radius $a(t)$ undergoing
deformations by the Ricci flow. The evolution truncates to a single
differential equation for $a(t)$ which is solved by
\be
a(t) = \sqrt{a^2(0) - (n-1)t} ~.
\ee
The solution exists for all time from $-\infty$ to $a^2(0)/(n-1)$,
until the space
fully collapses to a point for all $n \geq 2$. In the context of
$O(n+1)$ non-linear sigma models, it corresponds to the familiar
running of the coupling constant $g^2 = a^{-2}$,
\be
{1 \over {\tilde{g}}^2} = {1 \over g^2} + (n-1)
{\rm log}{\tilde{\Lambda} \over \Lambda}
\ee
with respect to the world-sheet renormalization scale $\Lambda^{-1}$,
setting $t = {\rm log} \Lambda^{-1}$. The field theory of a free compact
boson (Gaussian model) corresponds to $n=1$ and does not run. Sigma
models with more general target spaces of positive curvature
asymptote this special solution close to the big crunch singularity, since
they tend to become rounder and rounder as they shrink to a point.

Likewise, for the mean curvature flow, we consider the case of
concentric round $n$-spheres in $R^{n+1}$ with radius $a(t)$. Their
evolution reduces to a single differential equation for $a(t)$ that
yields
\be
a(t) = \sqrt{a^2(0) - 2n t} ~.
\ee
The solution exists for all time from $-\infty$ to $a^2(0)/2n$, until it
fully collapses to a point. The evolution of more general convex
hypersurfaces in $R^{n+1}$ asymptotes this special solution close
to the big crunch singularity. Note, however, that the contraction
rate of deforming $n$-spheres by mean curvature is always larger than
that of the Ricci flow. It is also non-zero for $n=1$, unlike the obvious
invariance of $S^1$
under the Ricci flow.
Thus, it is natural to expect that closed curves on positive curvature
spaces will also shrink faster than the ambient space, as they do in
most cases, when the combined
system of Ricci and mean curvature flows is under consideration.
However, there are examples of branes on curved spaces
where the configurations become singular
simultaneously.

Next, let us define {\em gradient solitons}
of the Ricci flow by the following
relation
\be
R_{\mu \nu} = -2 \nabla_{\mu} \nabla_{\nu} \Phi
\ee
for appropriately chosen field $\Phi(X)$. Solutions of this kind
arise when the metric deforms by pure diffeomorphisms with
generating vector field $\xi_{\mu} = - \partial_{\mu} \Phi$.
Gradient solitons can be regarded as fixed points of the
metric-dilaton flow and correspond to non-trivial conformal
field theories. The Euclidean black-hole in two dimensions with
metric and dilaton fields
\be
ds^2 = dr^2 + {\rm tanh}^2 r d \theta^2 ~, ~~~~~ \Phi (r) =
-{\rm log}({\rm cosh}r)
\ee
provides example of a rotationally symmetric gradient Ricci soliton
on $R^2$ with the
geometry of a semi-infinite long cigar, \cite{hami2}, \cite{witten}.

Likewise, gradient solitons of the mean curvature flow are defined
by equation
\be
\sum_{\sigma = 1}^{\rm codim}
H^{\sigma} {\hat{n}}_{\sigma}^{\mu} = - \partial^{\mu}
\Phi ~,
\ee
which describes deformations of branes by pure diffeomorphisms generated
by a vector field $\xi^{\mu} = - \partial^{\mu} \Phi$.
The two-dimensional plane supports translating solitons of the form
\be
y (x) = - {\rm log}({\rm cos} x)
\ee
with linear dilaton, $\Phi(y) = -y$, that does not affect the
conformal field theory of two free bosons on the bulk. This soliton is
a curve that asymptotes the lines $x = \pm \pi / 2$
as $y \rightarrow \infty$ and has a tip
located at the orgin of coordinates. It is has become known as
{\em grim-reaper} in the mathematics literature, \cite{grayson},
or {\em hair-pin}
in the physics literature, \cite{zamo2}.
Finally, there are simple solutions that arise by placing
mean curvature solitons on Ricci solitons, like a hair-pin on
a cigar.

As mentioned in the beginning,
supersymmetry puts severe constraints on the
geometry of the target space manifold and restricts it to be K\"ahler.
Thus, the natural question arises how the bulk renormalization group
equation combines with supersymmetry. Since supersymmetry
remains unbroken to
all orders in perturbation theory, the deformations of the metric
should preserve the K\"ahler class of the metric. Indeed, by introducing
a system of complex coordinates $(Z^a , {\bar{Z}}^a)$ on ${\cal M}$,
the renormalization of the metric follows the so called
Ricci-K\"ahler flow,
\be
{\partial \over \partial t} G_{a \bar{b}} = - R_{a \bar{b}}
\ee
with the required property.
This is a great simplification since all components of the metric
depend on a single scalar function, the K\"ahler potential, which
evolves accordingly, and the non-vanishing components of the
Ricci curvature tensor are simply expressed as
\be
R_{a \bar{b}} = - {\partial^2 \over \partial Z^a \partial {\bar{Z}}^b}
{\rm log} ({\rm det} G) ~.
\ee
On $S^2$ with K\"ahler metric
\be
ds^2 = 2 e^{\Phi (Z, \bar{Z}; t)} dZ d \bar{Z}
\ee
the Ricci flow takes the form
\be
{\partial \over \partial t} e^{\Phi} = {\partial^2 \Phi \over
\partial Z \partial \bar{Z}}
\ee
and can be (formally) integrated by group theoretical methods based on
B\"acklund transformations, \cite{bakas1}. Here, $\Phi$ should not be
confused with the dilaton field.

Extended supersymmetry puts more stringent constraints on the target
space geometry of sigma models, turning the K\"ahler condition into
hyper-K\"ahler, \cite{susy}. In this case there are three independent
K\"ahler structures and the metric is necessarily Ricci flat.
Thus, the models are conformal, since they correspond to fixed points
of the Ricci flow, and do not renormalize.
If, however, one is willing to drive them away
from the fixed points, by mathematical curiosity,
the Ricci-K\"ahler flow will only preserve
one K\"ahler structure and break the others. Conversely,
a whole sphere of K\"ahler structures will arise at the
fixed points of the flow on K\"ahler manifolds of dimension $4k$
and supersymmetry will be enhanced.

It should be noted at this point that Ricci flow is not the only
deformation of the metric compatible with supersymmetry.
Other geometric flows have been introduced on K\"ahler manifolds,
which also preserve the K\"ahler class of the metric, but they
lead to higher order non-linear differential equations. The prime
example is provided by the so called {\em Calabi flow} that is only
defined on K\"ahler manifolds and assumes the following form,
\cite{calabi},
\be
{\partial \over \partial t} G_{a \bar{b}} = {\partial^2 R \over
\partial Z^a \partial {\bar{Z}}^b}
\ee
in terms of the Ricci scalar curvature $R$. Clearly, it is
a fourth order equation, as opposed to the much simpler second
order Ricci-K\"ahler flow, and turns out to be much harder to
investigate by mini-superspace truncations. It preserves the
total volume of the space ${\cal M}$, unlike the Ricci flow, and
the fixed points, called {\em extremal metrics}, extremize the quadratic
curvature functional
\be
{\cal C} = \int_{\cal M} dV(G) ~ R^2 ~.
\ee
${\cal C}$ decreases monotonically along the flow, and, as such, it
acts as Lyapunov functional for the system. This is another
example of intrinsic curvature flows which is closely related
to the Ricci-K\"ahler flow by superevolution, \cite{bakas2}.

The Calabi flow has been used in mathematics as tool to study
a variety of non-linear problems in K\"ahler geometry, such as
uniformization, since extremal metrics are typically constant
curvature metrics. The round 2-sphere
is the most elementary example of this
kind. On the other hand, the type of deformations that are induced
on the metrics are quite different from the renormalization
of the bulk coupling of sigma models, and, as a result,
Calabi flow has not
yet found its proper place in quantum field theory. Of course,
this does not exclude the possibility to arrive at this flow in quantum
systems other than the sigma model. It is conceivable that the
renormalization group analysis of some (yet unknown) quantum field theory
may yield the combined system of second order equations
\be
{\partial \over \partial t} G_{a \bar{b}} = {\partial^2 \Psi \over
\partial Z^a \partial {\bar{Z}}^b} ~, ~~~~~
\Psi = - G^{c \bar{d}} {\partial^2 \over \partial Z^c \partial
{\bar{Z}}^d} {\rm log} ({\rm det} G)
\ee
when restricted on a certain line in the space of couplings
$(G, \Psi)$. Then, the Calabi flow for $G_{a \bar{b}}$ will simply
follow after elimination of $\Psi$.
Yet another possibility is that the Calabi flow
can be accommodated in the study of
transitions among different vacua of superstring theory, as
for the Ricci flow that accounts for the off-shell
description of tachyon condensation.
Although we will leave these questions open to future work, it is
worth mentioning here a concrete physical manifestation of the
two-dimensional Calabi flow in the classical theory of
general relativity.

Recall the theory of spherical gravitational waves in vacuum as
described by the class of four-dimensional Robinson-Trautman
radiative metrics, \cite{robi},
\be
ds^2 = 2r^2 e^{\Phi (Z, \bar{Z}; t)} dZ d \bar{Z} - 2dt dr
-H(Z, \bar{Z}, r, t) dt^2 ~.
\ee
Closed surfaces of constant $r$ and $t$ represent topological
2-spheres with complex coordinates $(Z, \bar{Z})$ and metric
coefficient given by $\Phi (Z, \bar{Z}; t)$. It can be verified that
some components
of Einstein's equations restrict the form of the function $H$ to
\be
H = r {\partial \Phi \over \partial t} - \Delta \Phi - {2m \over r} ~,
\ee
where $m$ is an integration constant with the interpretation of
mass parameter and
\be
\Delta = e^{-\Phi} \partial \bar{\partial}
\ee
is the Laplace-Beltrami operator on $S^2$.
When $m>0$, one may set without loss of generality
$3m = 1$ in appropriate units. Then, the remaining Einstein
equations yield
\be
\Delta \Delta \Phi + {\partial \Phi \over \partial t} = 0 ~,
\ee
which is known as Robinson-Trautman equation. It is identical to
the Calabi flow on $S^2$ with deformation parameter $t$ given by
the retarded time of the four-dimensional metric, \cite{tod}.
In this case
the extremal metric on $S^2$ is that of a round sphere, with
constant curvature, and the corresponding Robinson-Trautman metric
is no other but the static Schwarzschild solution of mass $m$ in
Eddington-Finkelstein frame. Time dependent solutions represent
purely outgoing gravitational radiation in space-time.

Finally, note that
the Calabi flow on $S^2$ can be (formally)
integrated by group theoretical methods,
\cite{bakas2}, as for the Ricci flow.
The integration of both Ricci and Calabi flows on two-dimensional
surfaces relies on their formulation as zero curvature conditions
$[\partial + A , \bar{\partial} + B] = 0$ with respect to $Z$ and
$\bar{Z}$. The deformation variable $t$ is fully encoded in the
structure of the algebra where the gauge connections
$A(Z, \bar{Z}; t)$ and $B(Z, \bar{Z}; t)$ take their values. The
relevant framework is provided by the class of infinite dimensional
{\em continual Lie algebras}, \cite{misha}, with appropriately
chosen Cartan operator. Thus, curvature dissipation becomes
internal affair of the symmetry group and does not contradict
integrability in two dimensions. Further details can be found in the
original papers \cite{bakas1}, \cite{bakas2}, where both
evolution equations are treated as continual Toda systems.

The last item that will be discussed here
is the so called {\em Willmore flow}. It describes deformations of
embedded branes in Riemannian geometry driven by their extrinsic
curvature, but it is fourth order equation as compared to the
second order mean curvature flow.
As such, it can be regarded as an extrinsic curvature analogue
of the Calabi flow. More explicitly, let us consider the
case of immersed submanifolds ${\cal N}$ in ${\cal M}$, which is
taken to be $R^n$ for simplicity, and
define the Willmore flow as gradient flow of the quadratic
curvature functional
\be
{\cal W} = {1 \over 2} \int_{\cal N} dV(g) ~ H^2 ~ ,
\ee
following, for example, \cite{will1}, \cite{will2},
and references therein.
$H^2$ is short-hand notation for the inner product of the
mean curvature vector with itself. ${\cal W}$ is
called Willmore functional although its roots date back to
the 18th century. One can add to ${\cal W}$
a multiple of the volume functional of the embedded
submanifold ${\cal N}$,
\be
S_{\rm ext} =
s_0 \int_{\cal N} dV(g) ~ H^2 ~ - t_0 \int_{\cal N} dV(g) ~ ,
\ee
with arbitrary coefficients $s_0$ and $t_0$,
and obtain interesting modifications of the evolution equation
as gradient flow; normalized evolutions can also be obtained in
this fashion.

${\cal W}$ provides the elastic energy
of curves, when the submanifold is one-dimensional, which by the
classic Bernoulli-Euler theory exhibits critical points (under
appropriate conditions) called {\em elasticae}; see, for
instance, \cite{will3} for a historic overview. For two-dimensional
submanifolds, the critical points of ${\cal W}$ are called
{\em Willmore surfaces};
see, for instance, \cite{will4} for a comprehensive account of
Willmore surfaces and related mathematical problems.
Since ${\cal W}$ is analogous to Calabi's
quadratic curvature functional ${\cal C}$ one may think of its
critical points as being extrinsic curvature analogues of
extremal metrics. Thus, one expects that the Willmore flow will
evolve a given submanifold towards one of the critical
points, modulo singularities.
Note, however, a difference with the Calabi flow which
does not derive from ${\cal C}$ as gradient flow; variants of the
Willmore flow can be introduced to match the difference, but they
are not be important here. Physical applications of fourth order
extrinsic curvature flows include interface diffusion.

In string theory there is already a well studied modification
of the classical Nambu-Goto action by adding quadratic extrinsic
curvature terms, \cite{will5}, as in the action $S_{\rm ext}$
above, thinking of ${\cal N}$ as two-dimensional world-sheet.
In this case strings have tension $t_0$ as well as rigidity
(or stiffness) $s_0$ that affects their behavior.
This is, of course, quite different from the interpretation
of $S_{\rm ext}$ as effective action for the beta functions
of deforming branes, which do not seem to receive higher order
curvature corrections of this type by 2-loop computations,
\cite{will6}. Branes do not become stiff within the quantum theory
of ordinary sigma models, and, as a result, the Willmore flow
can not be interpreted as renormalization group equation.
However, it might be interesting in this context to provide a
systematic definition of Dirichlet
sigma models for strings with rigidity and compute the effect that
the world-sheet extrinsic curvature terms may have on the beta
functions of embedded branes, as for the bulk.
The realization of Willmore flow
in quantum field theory remains an open problem.

In conclusion, we have considered the interplay of quantum
field theory and geometry as it manifests in the renormalization
group analysis of sigma models. Two-dimensional sigma models
with world-sheet boundary allow for bulk and boundary couplings
that depend on the energy scale. These theories have all the
necessary geometric data that allows to connect them with
the mathematical theory of geometric flows. Indeed,
the corresponding beta functions
realize and unite the Ricci and mean curvature flows, respectively,
and offer new possibilities in mathematics by also including
the effect of fluxes. It remains to be seen whether other
geometric evolution equations, driven by intrinsic or extrinsic
curvature terms, admit a similar realization in quantum
field theory. There is plenty of room for making new contributions
to this exciting area of current research that already had
great impact in science.

\vskip 1cm
\centerline{\bf Acknowledgements}

This work was supported in part by the
European Research and Training Network
``Constituents, Fundamental Forces and
Symmetries of the Universe" under contract
number MRTN-CT-2004-005104, the INTAS
programme ``Strings, Branes and Higher
Spin Fields" under contract number
03-51-6346 and the E$\Pi$AN programme
of the General Secretariat for Research
and Technology, Greece, under contract
number B.545. I am also grateful to the
conference organizers for their kind
invitation, financial support, and warm hospitality.


\begin{thebibliography}{3}
\bibitem{perel}
G. Perelman, ``The entropy formula for the Ricci flow and its
geometric applications", math.DG/0211159; ``Ricci flow with surgery
on three-manifolds", math.DG/0303109; ``Finite extinction time
for the solutions to the Ricci flow on certain three-manifolds",
math.DG/0307245.
\end{thebibliography}
\end{document}